\documentclass[preprint,showpacs,amsmath,amssymb]{revtex4-1}
\usepackage{amsmath}
\usepackage{amssymb}
\usepackage{graphicx}% Include figure files
\usepackage{dcolumn} % Align table columns on decimal point
\usepackage{bm} % bold math

\begin{document}

\title{Optical second harmonic generation from interfaces between heavy and ferromagnetic metals}

\author{T.V. Murzina}
\email{murzina@mail.ru}
\author{K.A. Lazareva}
\author{E.E. Shalygina}
\author{I.A. Kolmychek}
\affiliation{Department of Physics, Moscow State University, 119991 Moscow, Russia}
\author{E.A.Karashtin}
\author{N.S.Gusev}
\author{A.A.Fraerman}
\email{andr@ipmras.ru}
\affiliation{Institute for Physics of Microstructures RAS, Nizhny Novgorod 603950, GSP-105, Russia}

\date{\today}

\begin{abstract}
Surfaces and interfaces of magnetic nanostructures can reveal rather interesting and unusual properties that differ substantially from those of bulky materials. Here we apply the surface-sensitive method of optical second harmonic generation (SHG) for the studies of magnetization induced effects that appear in the nonlinear reflection from interfaces between ferromagnetic (Co) and heavy metals (Pt, Ta, W, Au, Ag, Cu). We demonstrate the appearance of magnetization induced variation in the p-polarized SHG intensity in the geometry of the longitudinal magneto-optical Kerr effect that is forbidden for homogeneous magnetic structures. This confirms the existence of chiral magnetic states at heavy metal/ferromagnet interfaces that appear due to the surface-induced Dzyaloshinskii-Moriya interaction. The related nonlinear chiroptical effect in the SHG intensity is proportional to the dc flexo-electric polarization that is shown to exist for chiral magnetic states at the considered interfaces.
\end{abstract}

%\pacs{07.78.+s, 07.60.-j, 75.50.Bb, 75.50.Cc, 75.75.Cd, 75.75.Fk, 78.66.-w}

\maketitle

\section{Introduction}
\begin{figure}[t]
\includegraphics[width=5.0in, keepaspectratio=true]{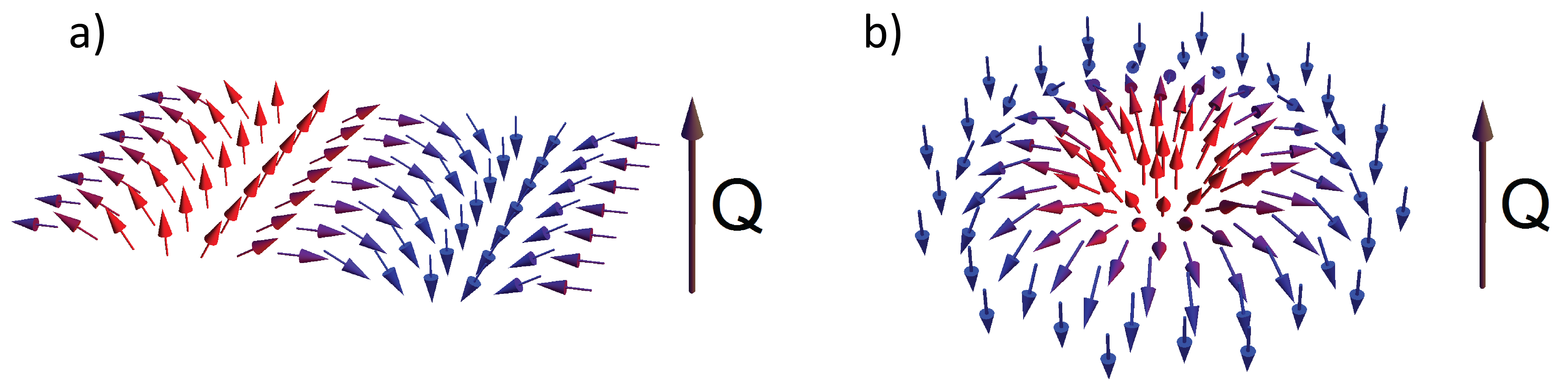}
\caption{\label{Fig_1} Magnetization distributions supporting the existence of non-zero flexo-magnetoelectric effect: (a) a magnetic cycloid and (b) a Neel-type skyrmion, which represents a part of a magnetic cycloid rolled up in a ring in the third dimension.}
\end{figure}

Studies of spin related effects at the interfaces between normal and ferromagnetic metals is a challenging task from the fundamental point of view, and are important for the further development of spintronic devices \cite{Zutic04}. Of particular interest is the study of the influence of spin - orbit interaction in a nonmagnetic metal with high values of spin-orbital constants on the magnetization distribution in a ferromagnetic metal. It was shown that this effect is due to the existence of an additional interaction of magnetic moments at the interface of a ferromagnetic and normal metals known as surface-induced Dzyaloshinskii-Moriya exchange interaction (DMI), which exists in noncentrosymmetric crystals and structures \cite{Kim13}. In the case of centrosymmetric metals, interface leads to the breaking of the inversion symmetry inherent to polycrystalline metals, including ferromagnetic ones. The energy of such interaction $\epsilon_{DM}$   can be written as:
\begin{equation} \label{Eq_1}
\epsilon_{DM}=-d (\mathbf{Q}\mathbf{n})
\end{equation}
where $d$ is a constant, $\mathbf{n}$ is a unit vector that determines the orientation of the interfaces, and  $\mathbf{Q}$ is a polar vector describing the flexo-magneto-electric effect \cite{Baryakhtar83, Katsura05, Mostovoy06} that exists in media with spatially-inhomogeneous distribution of magnetization, its $i-$th component is determined by the following expression:
\begin{equation} \label{Eq_2}
Q_{i}= q e_{ijk} [ \mathbf{m} \times \dfrac{\partial \mathbf{m}}{\partial x_j} ]_{k}
\end{equation}
Here $q$ is a constant, $e_{ijk}$ is the antisymmetric Levi-Civita tensor, $\mathbf{m}=\mathbf{M}/M$ is a unit magnetization vector. Dzyaloshinskii-Moriya interaction (DMI) described by (\ref{Eq_1}) leads to the formation of chiral magnetic distributions at the surface of a ferromagnet, such as magnetic cycloids and skyrmions \cite{Fert17}. Figure \ref{Fig_1} shows the magnetization distributions corresponding to the $“$Neel$”$ cycloid and  skyrmion. Importantly, both distributions and hence their arbitrary combinations are characterized by a polar vector $\mathbf{Q}$ which is the  flexoelectric polarization. Experimentally, the presence of surface-induced DMI can be revealed by inelastic light scattering involving spin excitations (Mandel'shtam-Brillouin scattering) and by spin-polarized tunneling microscopy \cite{Hellman17}.

On the other hand, optical second harmonic generation (SHG) is extremely sensitive to the state of the interface, which makes nonlinear magneto-optics an effective tool for studying the magnetic state of surfaces and interfaces \cite{Aktsipetrov11, Pan89}. It is known that second harmonic generation is forbidden for the bulk of centrosymmetric media, while at interfaces this symmetry is broken and the reflected SHG appears. Another mechanism for the SHG is the existence of the internal order parameter in a crystal, whose distribution lacks the inversion symmetry \cite{Landau8, Boyd08}. Such a situation is realized for example for magnetic systems characterized by the magnetic toroidal moment \cite{Krutyanskiy13, Kolmychek15}.

It was predicted recently that a new SHG mechanism appears for the media with chiral distribution of magnetization that is characterized by the polar vector $\mathbf{Q}$ \cite{Karashtin18}. The phenomenological expression for the macroscopic dipole moment induced at the double frequency can be written as:

\begin{equation} \label{Eq_3}
\begin{split}
P_{i}^{2\omega}= P_{0,i}^{2\omega}+P_{1M,i}^{2\omega}+P_{2M,i}^{2\omega}+P_{QM,i}^{2\omega}= \\
= P_{0,i}^{2\omega}+P_{1M,i}^{2\omega}+\Sigma \chi^{(2)}_{ijkl} E^{\omega}_j E^{\omega}_k Q_l + ...
\end{split}
\end{equation}
where the subscript in the SHG polarization vector corresponds to the ordinal number in the expansion in powers of M: the first term, $P_{0,i}^{2\omega}$, does not depend on the magnetization and the external magnetic field and corresponds to the so-called crystallographic contribution to nonlinear polarization. The second and third terms,  $P_{1M,i}^{2\omega}$ and $P_{2M,i}^{2\omega}$, are linear and quadratic in the magnetization of the layers, respectively. Finally, the last term in Eq. \ref{Eq_3} is due to the existence of the flexo-magnetoelectric polarization $\mathbb{Q}$. Formally, this term is of the second order with respect to M, but it contains its derivative. $E^{\omega}_j$ is the electric field strength of the fundamental wave. We do not take into account the $P_{2M,i}^{2\omega}$ term that arises due to second order of the uniform component of magnetization for fllowing reasons. First, we suppose that there is no average magnetization in the direction perpendicular to the applied external magnetic field (even in a zero external field there is no physical reason for this component to appear after magnetizing in a perpendicular direction). Next, the average of the magnetization along the applied field squared is even with respect to this field. We will later see that a new odd effect is observed due to the non-uniform magnetization. Hence this contribution to $P_{2M,i}^{2\omega}$ may also be neglected in the framework of current paper.

It is important to note that the NLO effects of the second-order in $M$ associated with the chiral interfactial magnetic state can be distinguished from the linear in magnetization effects by using the appropriate selection rules. If the magnetic moment of the sample is homogeneous and lies in the plane of incidence, then linear in magnetization dipole moment at the SHG wavelength is perpendicular to this plane regardless of the polarization of the pump wave \cite{Rzhevsky07}. Indeed, for a uniform and isotropic surface, we have
\begin{eqnarray} \label{Eq_4}
P_{1M,i}^{2\omega}&=& \chi_{isjkl} m_j n_s E_k^{\omega}E_l^{\omega};   \\ \nonumber
\chi_{isjkl} m_j n_s &=& \chi_1 \delta_{ki} e_{jsl} + \chi_2 \delta_{kl} e_{ijs} + \chi_3 \delta_{kj} e_{isl} + \chi_4 \delta_{sk} e_{ilj}
\end{eqnarray}
where $\delta_{ij}$ is the Kronecker symbol. Using (\ref{Eq_4}) it can be shown by direct constitution that the  induced dipole moment at the SHG wavelength, $\mathbf{P}_{1M}^{2\omega}$, is perpendicular to the plane of incidence of the fundamental radiation so that the generation of the p-polarized SHG is forbidden. This corresponds to the forbiddance of the magnetization-induced effect in p-polarized SHG under the longitudinal magnetization, which is consistent with the symmetry analysis performed previously \cite{Rzhevsky07}. At the same time, chiral distribution of magnetization that may exist at the interfaces should contribute to the p-polarized SHG through the term $\mathbf{P}_{QM}^{2\omega}$ from (\ref{Eq_3}). These SHG selection rules are used in this work for the visualization and studies of the chiral magnetic states at the interfaces between a normal and a ferromagnetic metals.

\section{Experimental section}
We studied a series of $Co(20nm)/Me(3nm)$ and $Me(3nm)/Co(2.5nm)/Me(3nm)$ samples, where the metals (\textit{Me} = Pt, Ta, W, Au, Ag, Cu) were chosen due to the available values of the spin-orbit interaction; the sequence of these metals corresponds to the decrease of the absolute value of the DM interaction at the Co / Me interface \cite{Fert17, Shahbazi18, Shahbazi18_2}. High vacuum magnetron sputtering was performed using AJA 2200 multichamber system onto amorphous $SiO_2$ or glass substrates at a basic pressure of 
$10^{-5}$ Pa. The deposited metal films had a polycrystalline structure with the grain sizes of about 10 nm.

\begin{figure}[t]
\includegraphics[width=4.25in, keepaspectratio=true]{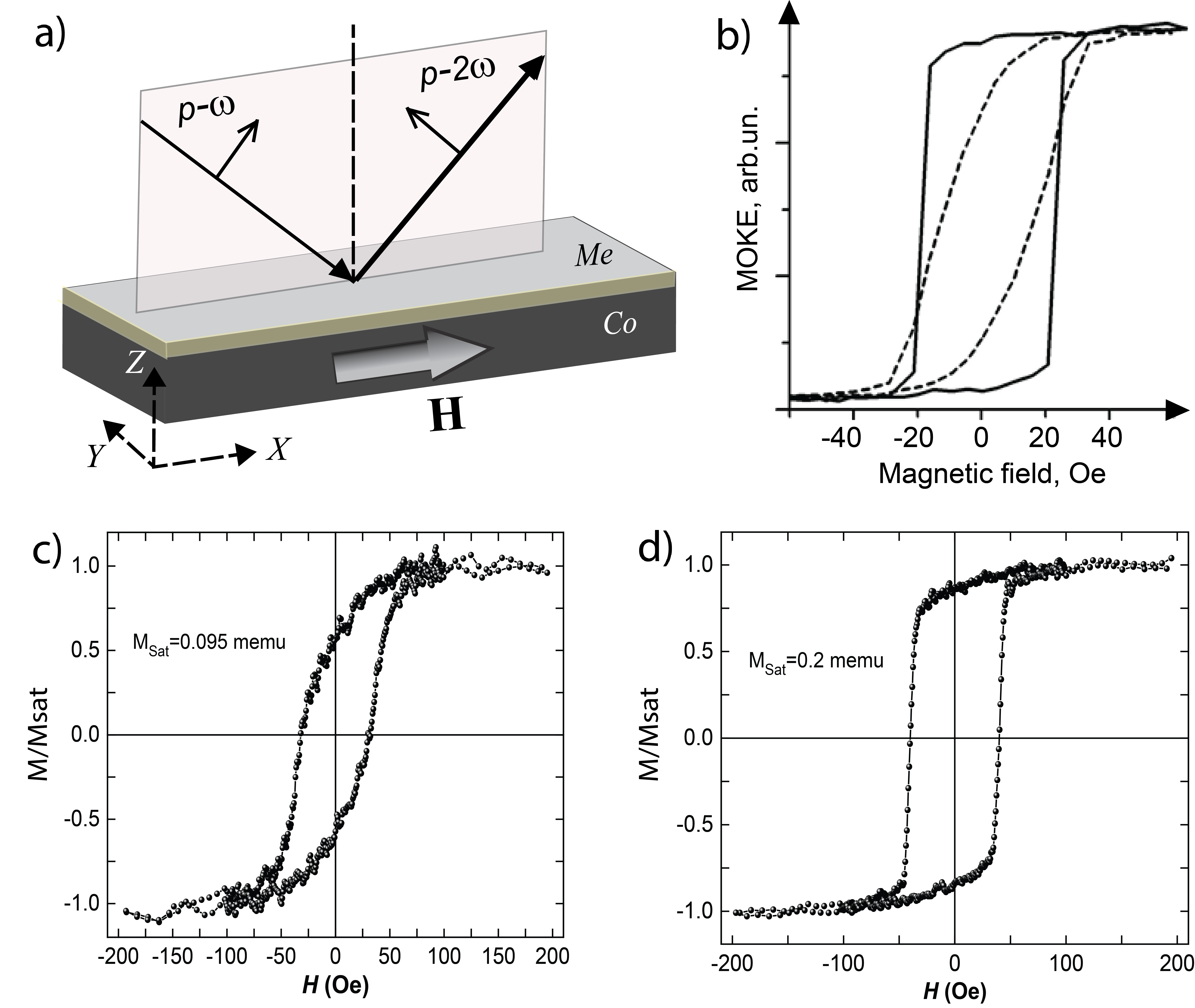} 
%\caption{\label{Fig_2} (a) Schematic view of the experimental geometry. (b) Linear MOKE hysteresis loops for the two orthogonal azimuthal orientations of the sample in the laboratory set-up. (с),(d) - magnetization of the Pt/Co/Pt structure for the same orientations of the structure.}
%\end{figure}
\caption{\label{Fig_2} (a) Schematic view of the experimental geometry. (b) Linear MOKE hysteresis loops for the orthogonal azimuthal orientations of the sample in the laboratory set-up. (c),(d) magnetization dependencies of the Pt/Co/Pt structure for the same azimuthal positions of the samples.}
\end{figure}

Magnetic properties of the samples were studied by the vibrational magnetometry and magneto-optical (MO) methods in the geometry of the longitudinal MO Kerr effect (MOKE). In the MO measurements, we studied the rotation of the polarization plane of the HeNe laser beam radiation ($\lambda$ = 632 nm) reflected from the surface of the structure depending on the magnitude of the applied dc magnetic field. Figure 2,b shows the MOKE hysteresis loops for a Co(20 nm)/Pt(3nm) sample, which are typical for all samples of this series. The two curves correspond to the orthogonal azimuthal orientations of the sample placed between the poles of the electromagnet. Pronounced differences in the shape of these hysteresises confirm the magnetic anisotropy of the composed films, similar to known for thin films of transition metals \cite{Wilts68, Andr60}. Below we present the results of the SHG experiments for the external magnetic field applied along the axis that corresponds to solid line in Figure~\ref{Fig_2},b (close to the easy magnetization axis).

For the SHG experiments linearly polarized radiation of a Ti:sapphire laser was used at a wavelength of 780 nm, a pulse duration of 80 fs, a repetition frequency of 80 MHz, and a mean power of 100 mW. The p-polarized SHG radiation, reflected from the film, was spectrally selected by BG39 Schott color filters and detected by a photomultiplier operating in the photon counting mode.

In order to visualize the effect of chiral magnetic distribution at the interface of a heavy metal and cobalt, we performed the measurements of the  p-polarized SHG intensity dependence of the applied longitudinal magnetic field. 
As noted above, linear in magnetization modulation of the p-polarized SHG intensity in this case should be absent for the structures with a uniform distribution of magnetization, and the only possible mechanism of this effect is the existence of chiral magnetic states with induced polar $\mathbb{Q}$ vector at the interface. For comparison, ``allowed'' nonlinear MOKE measurements at the SHG wavelength were performed as the analyzer prior to the detection system was oriented at 45$^{\circ}$ with respect to the p-polarization, so that the magnetic field induced rotation of the SHG polarization plane was measured. 

\section{Experimental results}
Figure~\ref{Fig_2},a shows the scheme of the experiment for the studies of the $Q-$induced SHG contribution, when a sample is placed in the longitudinal magnetic field and the p-polarized SHG induced by p-polarized fundamental radiation is studied. Panels \textit{c} and \textit{d} confirm the magnetic anisotropy of the samples.
\begin{figure}[t]
\includegraphics[width=4.25in, keepaspectratio=true]{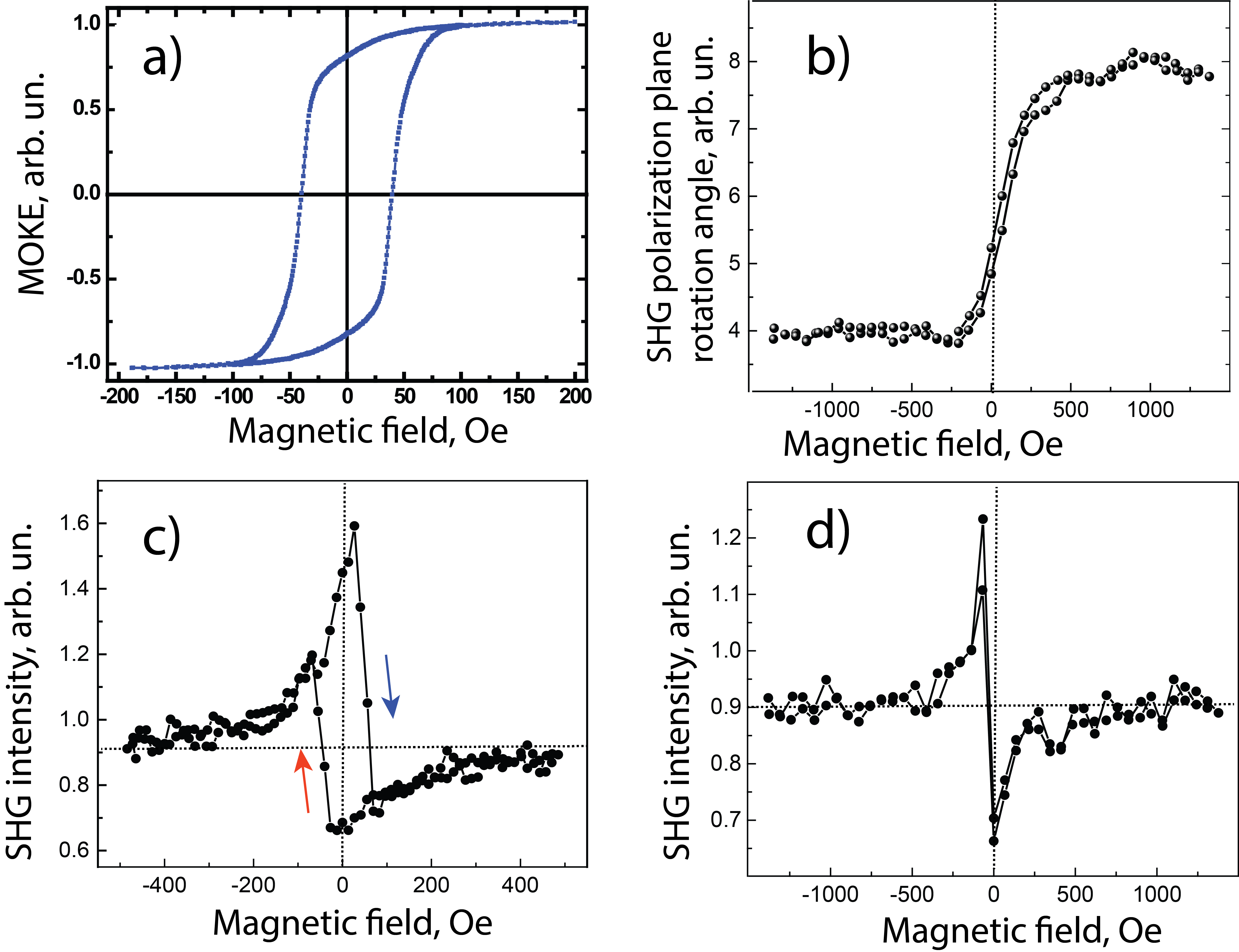}%
\caption{\label{Fig_3} Linear and nonlinear MOKE measurements for Co(20 nm)/Pt(3 nm) structure: (a) MOKE hysteresis loop, (b) longitudinal MOKE in SHG; (c,d) magnetic field induced dependence of the p-polarized SHG intensity for different ranges of the longitudinal dc magnetic field. Inset in panel (d) shows the dependence of the SHG magnetic contrast on the magnetic field.}
\end{figure}
First of all we searched for the composition of ferromagnetic and heavy metals that reveal the magnetic field induced effect for the ``forbidden'' polarization combination discussed above, that is p-polarized SHG and longitudinal magnetization. It turned out that this effect exists for the structures that contain Pt or Ta nanolayers shown in Figures \ref{Fig_3} and \ref{Fig_4}, while for other nonmagnetic metals neighbor to Co films it is absent within the experimental error. This supports the idea of the important role of surface-induced DMI that has been actively studied for  the interfaces of heavy and ferromagnetic metals such as Co/Pt and Co/Ta ones \cite{Belmeguenai15, Woo14}. 

Figure~\ref{Fig_3} accumulates the results obtained for the Co(20 nm)/Pt(3 nm) bilayer structure with a single magnetic interface. Linear MOKE  measurements shown in panel \textit{a}  allow to estimate the saturating field that is less than 100 Oe and coercitivity of approximately 50 Oe, which characterize primarily the magnetic properties of the bulky cobalt film. The comparison with the SHG hysteresis loop obtained for the longitudinal nonlinear MOKE geometry (see Figure \ref{Fig_3},b) shows that in the nonlinear-optical case the saturating magnetic field is a few times larger and is about 0.5 kOe. This difference should be attributed to the surface nature of the  SHG process and indicates that the magnetic properties of the Co/Pt interface differ substantially from those of bulky Co film. 
One more issue to be underlined is the large value of the nonlinear MO effect as compared to linear-optical one; the magnetization-induced SHG polarization plane rotation is about tens of degrees, which exceeds the MOKE by more than an order of magnitude. 
\begin{figure}[t]
\includegraphics[width=4.25in, keepaspectratio=true]{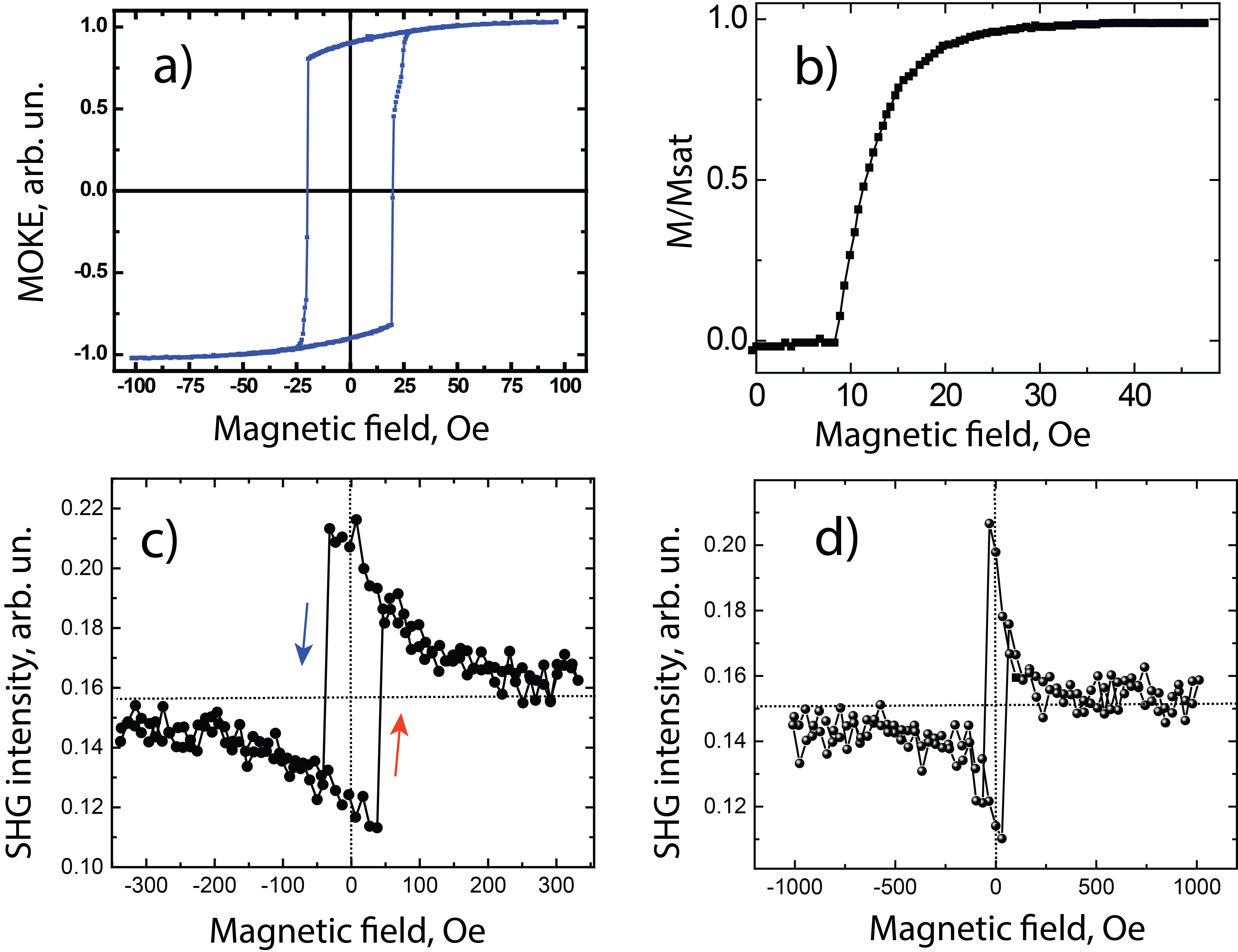}%
\caption{\label{Fig_4} Linear and nonlinear MOKE measurements for Co(20 nm)/Ta(3 nm) structure. (a) MOKE hysteresis loop; (b) magnetization dependence obtained from vibrating-sample magnetometer; (c,d) magnetic field induced dependence of the p-polarized SHG intensity for different ranges of the longitudinal dc magnetic field. Inset in panel (d) shows the dependence of the SHG magnetic contrast on the magnetic field.}
\end{figure}

Magnetic hysteresis loops in the SHG intensity for the Co/Pt bilayer film for the p-polarized SHG are shown in Fig. \ref{Fig_3},c,d. One can see a clear modulation of the SHG intensity, which is evident for the magnetic field up to 400-500 Oe; for larger field strength it decreases continuously down to zero (within the experimental accuracy). Importantly, this loop demonstrates SHG hysteresis of approximately 100 Oe in width, in consistence with the MOKE data. For the characterization of the SHG intensity modulation with the applied magnetic field, SHG magnetic contrast is commonly introduced as
$
\rho_{H}^{2\omega}= \dfrac{I_{+H}^{2\omega}-I_{-H}^{2\omega}}{I_{+H}^{2\omega}+I_{-H}^{2\omega}}
$
where $I_{+H}^{2\omega}$ and $I_{-H}^{2\omega}$ stand for the SHG intensity measured for the opposite orientations of the magnetic field $H$. In terms of $
\rho_{H}^{2\omega}$, SHG hysteresis loops for Co/Pt bilayer film are characterized by negative values of the SHG magnetic contrast for positive $H$ and vise versa. The sign of the $\rho_{H}^{2\omega}(Co/Pt)$ will be compared below with that of the $Co/Ta$ bilayer. Importantly, the positive orientation of the magnetic field is chosen on probation, still it is kept unchanged for all the measurements discussed in the paper.

Figure \ref{Fig_4} shows analogous results for the Co(20 nm)/Ta(3 nm) structure. It can be seen from panel~b) that the saturation of magnetization of the Co film checked by vibrating-sample magnetometer takes place at 25-30 Oe. This corresponds to the linear MOKE hysteresis (panel~a). Similarly to the Co/Pt bilayer, a nice SHG hysteresis loop is observed in the rotation of the SHG polarization plane. Moreover, the ``forbidden'' intensity effect for the p-polarized SHG also exists (Fig. \ref{Fig_4},c,d), while the crawl direction of the hysteresis loop is inversed as compared to the Co/Pt bilayer film. Consequently, the sign of $\rho_{H}^{2\omega}$ is also changed so that the positive values of the SHG magnetic contrast are attained for positive $H$ and vise versa.  

\begin{figure}[t]
\includegraphics[width=4.25in, keepaspectratio=true]{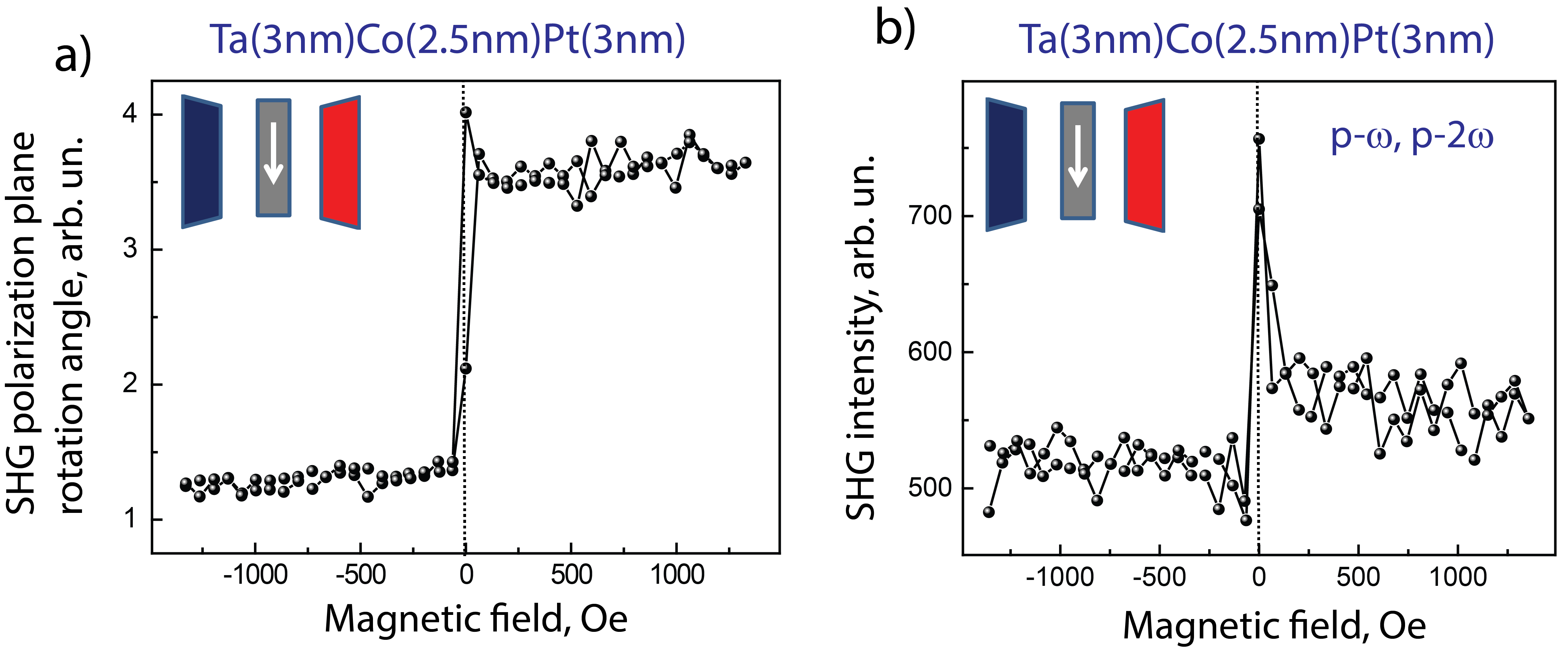}%
\caption{\label{Fig_5} Nonlinear MOKE measurements for Ta(3 nm)/Co(2.5 nm)/Pt(3 nm) structure: (a) rotation of the SHG polarization plane for the longitudinal MOKE geometry; (b) magnetic field induced dependence of the p-polarized SHG intensity.}
\end{figure}

Having in mind that the sign of the ``forbidden'' SHG intensity effect is different for the Co/Pt and Co/Ta interfaces, we composed the asymmetric hybrid Ta/Co/Pt structure with the thicknesses of all the layers of 3 nm. Figure~\ref{Fig_5} contains the results of the SHG measurements for two different orientations of the sample with respect to the external magnetic field and the plane of light incidence. Panel a) corresponds to the nonlinear MOKE and show pronounced SHG polarization plane rotation  with different saturation magnetic field values, which indicate the expected magnetic anisotropy of the structure. %Besides, SHG behaviour close to $H=0$ is quite different for various orientation of the easy magnetization axis with respect to the applied magnetic field, and should be attributed to a complicated display of the magnetization reversal in the nonlinear response of the structure. 
%
%Indeed, we have polycrystalline films grown by the dc magnetron sputtering. Magnetic anisotropy of the Co film is connected to this polycrystalline structure which also plays an important role in the process of magnetization reveresal of the sample \cite{Gusev13}. Therefore one may expect that the magnetic SHG response would be different for two different orientations of sample in the fields at which it is remagnetized.
%
Panel b) shows the magnetic field induced dependencies of the p-polarized SHG intensity for different ranges of the longitudinal magnetic field. An important issue here is that the non-zero SHG magnetic contrast remains even in large magnetic field up to 1.5 kOe. The latter result differs substantially the 3-layered Ta/Co/Pt nanostructure from bilayer samples with 20 nm thick Co films described above. 

\begin{figure}[t]
\includegraphics[width=4.25in, keepaspectratio=true]{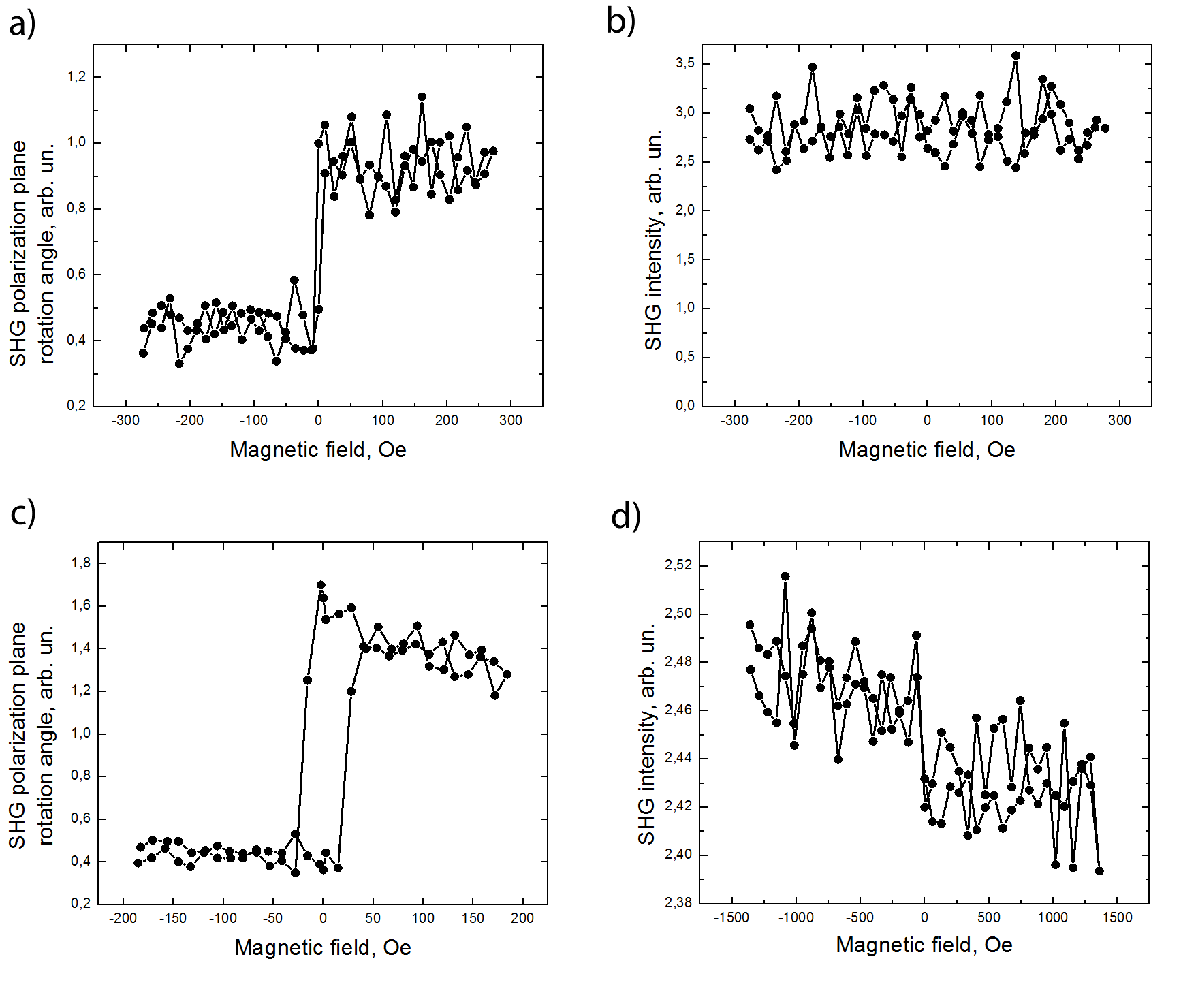}%
\caption{\label{Fig_6} Nonlinear MOKE measurements for Ta(3 nm)/Co(2.5 nm)/Ta(3 nm) (a,b) and Pt(3 nm)/Co(2.5 nm)/Pt(3 nm) (c,d) structure: (a,c) rotation of the SHG polarization plane for the longitudinal MOKE geometry; (b,d) magnetic field induced dependence of the p-polarized SHG intensity.}
\end{figure}

We have performed the same measurements for a symmetric Ta/Co/Ta multilayer sample (Figure~\ref{Fig_6},a,b). This sample does not demonstrate the ``forbidden'' SHG effect (Figure~\ref{Fig_6},b). However it is clearly seen in Figure~\ref{Fig_6},a that the SHG plane rotation exists for the sample. This may be attributed to the absorption of light in the Co layer which makes two Co/Ta interfaces not identical. The same measurements were carried out for a Pt/Co/Pt sample (see Figure~\ref{Fig_6},c,d). This semple demonstrated a much weaker ``forbidden'' effect than that of the  Ta/Co/Pt sample. Indeed, it can be estimated from Figure~\ref{Fig_6},d that the relative value of the effect is approximately $1.5\%$. At the same time, Figure~\ref{Fig_5},b shows that the ``forbidden'' effect for the Ta/Co/Pt sample is greater than $6\%$.

\section{Discussion}
When discussing the observed effects in SHG that are forbidden for homogeneously magnetized structures it is worth realizing the space localization of different SHG sources. It is commonly assumed that interface induced  breaking of the inversion symmetry leads to the appearance of surface dipole susceptibility localized in approximately 1 nm thick interface layer, which can posses magnetic properties different from bulky ones. Besides, electric quadruple SHG source originates from the skin layer that is about 20 nm for the visible spectral range in the case of metals. Thus both the interfaces and bulky Co participated in the SHG process for Co/Ta and Co/Pt bilayers, so that the symmetry allowed SHG Kerr effect qualitatively correlates with the linear MOKE measurements as they are mostly determined by the magnetic properties of bulk Co films. (However, saturation may occur at different fields for linear and SHG effects which we observed for a Co/Pt bilayer. This is definitely attributed to the strong role of the surface in SHG.)

At the same time, for the p-polarized SHG and longitudinal magnetization the contribution from the bulk of the film is not expected, so the interfaces should govern the magnetic effects in SHG. 
We show experimentally that in this ``forbidden'' SHG geometry nonzero magnetization-induced SHG appears only for Co/Ta and Co/Pt interfaces, which  confirm our hypothesis about the relationship between this effect and the magnitude of the surface-induced DM interaction at the interface between a ferromagnet and a heavy metal, which was actively studied for the Co/Pt and Co/Ta interfaces \cite{Belmeguenai15, Woo14}. This interaction is known to have different sign for these interfaces, which also correlates with our SHG studies which reveal opposite signs of the SHG magnetic contrast (i.e. of magnetic field induced effect in the SHG intensity) for these structures. 

It is known that if two interfaces of a ferromagnetic layer have different signs of DM interaction this makes the chiral magnetic structures easier to appear \cite{Min18}. So we expected that the ``forbidden'' effect in the intensity of p-polarized SHG would increase in Ta/Co/Pt layered structure, which combines the two interfaces with the opposite directions of the normal to the Co layer and of the DMI constant. This explicitly is seen in the experiment (Figure \ref{Fig_5}).  Moreover, one can suppose that the interaction of the two closely situated interfaces with chiral magnetization distribution is rather strong, so that a non-zero SHG magnetic contrast remains for the whole range of the longitudinal magnetic field under study.

We have also checked the ``forbidden'' effect for samples that consist of a Co layer surrounded by two layers made of the same material (Ta or Pt). In this case the effect is zero (for Ta/Co/Ta) or close to zero (for Pt/Co/Pt). This nonzero effect is possibly caused by slightly different properties of Pt grown on a substrate and on a Co layer which is not important for Ta known as a good material to remove strain in the sample. However it is important that if two similar heavy metals are chosen the effect tends to zero.

\section{Summary}
Summing up, discussed experiments confirm the possibility for the visualization of magnetic chiral states at the interface normal metal/ferromagnet by means of the SHG-based nonlinear magneto-optical technique. We demonstrate the appearance of the SHG magnetic field induced changes for the geometry of the experiments that excludes the observation of common (allowed) nonlinear magneto-optical Kerr effect. These variations are associated with the magnetic chirality-driven SHG contribution at the normal metal/ferromagnet interface. The occurrence of chirality in the magnetization distribution of a ferromagnet interface with a SOC metal is due to the surface-induced Dzyaloshinskii-Moriyla interaction.

\section{Acknowledgements}
This work was supported by the Russian Science Foundation (Grant No. 16-12-10340).

\bibliography{ArxiV}

\providecommand{\noopsort}[1]{}\providecommand{\singleletter}[1]{#1}%
\begin{thebibliography}{22}%
\makeatletter
\providecommand \@ifxundefined [1]{%
 \@ifx{#1\undefined}
}%
\providecommand \@ifnum [1]{%
 \ifnum #1\expandafter \@firstoftwo
 \else \expandafter \@secondoftwo
 \fi
}%
\providecommand \@ifx [1]{%
 \ifx #1\expandafter \@firstoftwo
 \else \expandafter \@secondoftwo
 \fi
}%
\providecommand \natexlab [1]{#1}%
\providecommand \enquote  [1]{``#1''}%
\providecommand \bibnamefont  [1]{#1}%
\providecommand \bibfnamefont [1]{#1}%
\providecommand \citenamefont [1]{#1}%
\providecommand \href@noop [0]{\@secondoftwo}%
\providecommand \href [0]{\begingroup \@sanitize@url \@href}%
\providecommand \@href[1]{\@@startlink{#1}\@@href}%
\providecommand \@@href[1]{\endgroup#1\@@endlink}%
\providecommand \@sanitize@url [0]{\catcode `\\12\catcode `\$12\catcode
  `\&12\catcode `\#12\catcode `\^12\catcode `\_12\catcode `\%12\relax}%
\providecommand \@@startlink[1]{}%
\providecommand \@@endlink[0]{}%
\providecommand \url  [0]{\begingroup\@sanitize@url \@url }%
\providecommand \@url [1]{\endgroup\@href {#1}{\urlprefix }}%
\providecommand \urlprefix  [0]{URL }%
\providecommand \Eprint [0]{\href }%
\providecommand \doibase [0]{http://dx.doi.org/}%
\providecommand \selectlanguage [0]{\@gobble}%
\providecommand \bibinfo  [0]{\@secondoftwo}%
\providecommand \bibfield  [0]{\@secondoftwo}%
\providecommand \translation [1]{[#1]}%
\providecommand \BibitemOpen [0]{}%
\providecommand \bibitemStop [0]{}%
\providecommand \bibitemNoStop [0]{.\EOS\space}%
\providecommand \EOS [0]{\spacefactor3000\relax}%
\providecommand \BibitemShut  [1]{\csname bibitem#1\endcsname}%
\let\auto@bib@innerbib\@empty
%</preamble>
\bibitem [{\citenamefont {\ifmmode \check{Z}\else
  \v{Z}\fi{}uti\ifmmode~\acute{c}\else \'{c}\fi{}}\ \emph
  {et~al.}(2004)\citenamefont {\ifmmode \check{Z}\else
  \v{Z}\fi{}uti\ifmmode~\acute{c}\else \'{c}\fi{}}, \citenamefont {Fabian},\
  and\ \citenamefont {Das~Sarma}}]{Zutic04}%
  \BibitemOpen
  \bibfield  {author} {\bibinfo {author} {\bibfnamefont {I.}~\bibnamefont
  {\ifmmode \check{Z}\else \v{Z}\fi{}uti\ifmmode~\acute{c}\else \'{c}\fi{}}},
  \bibinfo {author} {\bibfnamefont {J.}~\bibnamefont {Fabian}}, \ and\ \bibinfo
  {author} {\bibfnamefont {S.}~\bibnamefont {Das~Sarma}},\ }\href {\doibase
  10.1103/RevModPhys.76.323} {\bibfield  {journal} {\bibinfo  {journal} {Rev.
  Mod. Phys.}\ }\textbf {\bibinfo {volume} {76}},\ \bibinfo {pages} {323}
  (\bibinfo {year} {2004})}\BibitemShut {NoStop}%
\bibitem [{\citenamefont {Kim}\ \emph {et~al.}(2013)\citenamefont {Kim},
  \citenamefont {Lee}, \citenamefont {Lee},\ and\ \citenamefont
  {Stiles}}]{Kim13}%
  \BibitemOpen
  \bibfield  {author} {\bibinfo {author} {\bibfnamefont {K.-W.}\ \bibnamefont
  {Kim}}, \bibinfo {author} {\bibfnamefont {H.-W.}\ \bibnamefont {Lee}},
  \bibinfo {author} {\bibfnamefont {K.-J.}\ \bibnamefont {Lee}}, \ and\
  \bibinfo {author} {\bibfnamefont {M.~D.}\ \bibnamefont {Stiles}},\ }\href
  {\doibase 10.1103/PhysRevLett.111.216601} {\bibfield  {journal} {\bibinfo
  {journal} {Phys. Rev. Lett.}\ }\textbf {\bibinfo {volume} {111}},\ \bibinfo
  {pages} {216601} (\bibinfo {year} {2013})}\BibitemShut {NoStop}%
\bibitem [{\citenamefont {Bar'yakhtar}\ \emph {et~al.}(1983)\citenamefont
  {Bar'yakhtar}, \citenamefont {L'vov},\ and\ \citenamefont
  {Yablonskii}}]{Baryakhtar83}%
  \BibitemOpen
  \bibfield  {author} {\bibinfo {author} {\bibfnamefont {V.~G.}\ \bibnamefont
  {Bar'yakhtar}}, \bibinfo {author} {\bibfnamefont {V.~A.}\ \bibnamefont
  {L'vov}}, \ and\ \bibinfo {author} {\bibfnamefont {D.~A.}\ \bibnamefont
  {Yablonskii}},\ }\href@noop {} {\bibfield  {journal} {\bibinfo  {journal}
  {Pis'ma Zh. Eksp. Teor. Fiz.}\ }\textbf {\bibinfo {volume} {37}},\ \bibinfo
  {pages} {565} (\bibinfo {year} {1983})}\BibitemShut {NoStop}%
\bibitem [{\citenamefont {Katsura}\ \emph {et~al.}(2005)\citenamefont
  {Katsura}, \citenamefont {Nagaosa},\ and\ \citenamefont
  {Balatsky}}]{Katsura05}%
  \BibitemOpen
  \bibfield  {author} {\bibinfo {author} {\bibfnamefont {H.}~\bibnamefont
  {Katsura}}, \bibinfo {author} {\bibfnamefont {N.}~\bibnamefont {Nagaosa}}, \
  and\ \bibinfo {author} {\bibfnamefont {A.~V.}\ \bibnamefont {Balatsky}},\
  }\href {\doibase 10.1103/PhysRevLett.95.057205} {\bibfield  {journal}
  {\bibinfo  {journal} {Phys. Rev. Lett.}\ }\textbf {\bibinfo {volume} {95}},\
  \bibinfo {pages} {057205} (\bibinfo {year} {2005})}\BibitemShut {NoStop}%
\bibitem [{\citenamefont {Mostovoy}(2006)}]{Mostovoy06}%
  \BibitemOpen
  \bibfield  {author} {\bibinfo {author} {\bibfnamefont {M.}~\bibnamefont
  {Mostovoy}},\ }\href {\doibase 10.1103/PhysRevLett.96.067601} {\bibfield
  {journal} {\bibinfo  {journal} {Phys. Rev. Lett.}\ }\textbf {\bibinfo
  {volume} {96}},\ \bibinfo {pages} {067601} (\bibinfo {year}
  {2006})}\BibitemShut {NoStop}%
\bibitem [{\citenamefont {Fert}\ \emph {et~al.}(2017)\citenamefont {Fert},
  \citenamefont {Reyren},\ and\ \citenamefont {Cros}}]{Fert17}%
  \BibitemOpen
  \bibfield  {author} {\bibinfo {author} {\bibfnamefont {A.}~\bibnamefont
  {Fert}}, \bibinfo {author} {\bibfnamefont {N.}~\bibnamefont {Reyren}}, \ and\
  \bibinfo {author} {\bibfnamefont {V.}~\bibnamefont {Cros}},\ }\href@noop {}
  {\bibfield  {journal} {\bibinfo  {journal} {Nature Reviews Materials}\
  }\textbf {\bibinfo {volume} {2}},\ \bibinfo {pages} {17031} (\bibinfo {year}
  {2017})}\BibitemShut {NoStop}%
\bibitem [{\citenamefont {Hellman}\ \emph {et~al.}(2017)\citenamefont
  {Hellman}, \citenamefont {Hoffmann}, \citenamefont {Tserkovnyak},
  \citenamefont {Beach}, \citenamefont {Fullerton}, \citenamefont {Leighton},
  \citenamefont {MacDonald}, \citenamefont {Ralph}, \citenamefont {Arena},
  \citenamefont {D\"urr}, \citenamefont {Fischer}, \citenamefont {Grollier},
  \citenamefont {Heremans}, \citenamefont {Jungwirth}, \citenamefont {Kimel},
  \citenamefont {Koopmans}, \citenamefont {Krivorotov}, \citenamefont {May},
  \citenamefont {Petford-Long}, \citenamefont {Rondinelli}, \citenamefont
  {Samarth}, \citenamefont {Schuller}, \citenamefont {Slavin}, \citenamefont
  {Stiles}, \citenamefont {Tchernyshyov}, \citenamefont {Thiaville},\ and\
  \citenamefont {Zink}}]{Hellman17}%
  \BibitemOpen
  \bibfield  {author} {\bibinfo {author} {\bibfnamefont {F.}~\bibnamefont
  {Hellman}}, \bibinfo {author} {\bibfnamefont {A.}~\bibnamefont {Hoffmann}},
  \bibinfo {author} {\bibfnamefont {Y.}~\bibnamefont {Tserkovnyak}}, \bibinfo
  {author} {\bibfnamefont {G.~S.~D.}\ \bibnamefont {Beach}}, \bibinfo {author}
  {\bibfnamefont {E.~E.}\ \bibnamefont {Fullerton}}, \bibinfo {author}
  {\bibfnamefont {C.}~\bibnamefont {Leighton}}, \bibinfo {author}
  {\bibfnamefont {A.~H.}\ \bibnamefont {MacDonald}}, \bibinfo {author}
  {\bibfnamefont {D.~C.}\ \bibnamefont {Ralph}}, \bibinfo {author}
  {\bibfnamefont {D.~A.}\ \bibnamefont {Arena}}, \bibinfo {author}
  {\bibfnamefont {H.~A.}\ \bibnamefont {D\"urr}}, \bibinfo {author}
  {\bibfnamefont {P.}~\bibnamefont {Fischer}}, \bibinfo {author} {\bibfnamefont
  {J.}~\bibnamefont {Grollier}}, \bibinfo {author} {\bibfnamefont {J.~P.}\
  \bibnamefont {Heremans}}, \bibinfo {author} {\bibfnamefont {T.}~\bibnamefont
  {Jungwirth}}, \bibinfo {author} {\bibfnamefont {A.~V.}\ \bibnamefont
  {Kimel}}, \bibinfo {author} {\bibfnamefont {B.}~\bibnamefont {Koopmans}},
  \bibinfo {author} {\bibfnamefont {I.~N.}\ \bibnamefont {Krivorotov}},
  \bibinfo {author} {\bibfnamefont {S.~J.}\ \bibnamefont {May}}, \bibinfo
  {author} {\bibfnamefont {A.~K.}\ \bibnamefont {Petford-Long}}, \bibinfo
  {author} {\bibfnamefont {J.~M.}\ \bibnamefont {Rondinelli}}, \bibinfo
  {author} {\bibfnamefont {N.}~\bibnamefont {Samarth}}, \bibinfo {author}
  {\bibfnamefont {I.~K.}\ \bibnamefont {Schuller}}, \bibinfo {author}
  {\bibfnamefont {A.~N.}\ \bibnamefont {Slavin}}, \bibinfo {author}
  {\bibfnamefont {M.~D.}\ \bibnamefont {Stiles}}, \bibinfo {author}
  {\bibfnamefont {O.}~\bibnamefont {Tchernyshyov}}, \bibinfo {author}
  {\bibfnamefont {A.}~\bibnamefont {Thiaville}}, \ and\ \bibinfo {author}
  {\bibfnamefont {B.~L.}\ \bibnamefont {Zink}},\ }\href {\doibase
  10.1103/RevModPhys.89.025006} {\bibfield  {journal} {\bibinfo  {journal}
  {Rev. Mod. Phys.}\ }\textbf {\bibinfo {volume} {89}},\ \bibinfo {pages}
  {025006} (\bibinfo {year} {2017})}\BibitemShut {NoStop}%
\bibitem [{\citenamefont {Aktsipetrov}(2011)}]{Aktsipetrov11}%
  \BibitemOpen
  \bibfield  {author} {\bibinfo {author} {\bibfnamefont {O.~A.}\ \bibnamefont
  {Aktsipetrov}},\ }\href@noop {} {\bibfield  {journal} {\bibinfo  {journal}
  {J. Opt. Soc. Am. B}\ }\textbf {\bibinfo {volume} {32}},\ \bibinfo {pages}
  {A27} (\bibinfo {year} {2011})}\BibitemShut {NoStop}%
\bibitem [{\citenamefont {Pan}\ \emph {et~al.}(1989)\citenamefont {Pan},
  \citenamefont {Wei},\ and\ \citenamefont {Shen}}]{Pan89}%
  \BibitemOpen
  \bibfield  {author} {\bibinfo {author} {\bibfnamefont {R.-P.}\ \bibnamefont
  {Pan}}, \bibinfo {author} {\bibfnamefont {H.~D.}\ \bibnamefont {Wei}}, \ and\
  \bibinfo {author} {\bibfnamefont {Y.~R.}\ \bibnamefont {Shen}},\ }\href
  {\doibase 10.1103/PhysRevB.39.1229} {\bibfield  {journal} {\bibinfo
  {journal} {Phys. Rev. B}\ }\textbf {\bibinfo {volume} {39}},\ \bibinfo
  {pages} {1229} (\bibinfo {year} {1989})}\BibitemShut {NoStop}%
\bibitem [{\citenamefont {Landau}\ and\ \citenamefont
  {Lifshitz}(1984)}]{Landau8}%
  \BibitemOpen
  \bibfield  {author} {\bibinfo {author} {\bibfnamefont {L.~D.}\ \bibnamefont
  {Landau}}\ and\ \bibinfo {author} {\bibfnamefont {E.~M.}\ \bibnamefont
  {Lifshitz}},\ }\href@noop {} {\emph {\bibinfo {title} {Course of Theoretical
  Physics, Vol. 8: Electrodynamics of Continuous Media}}}\ (\bibinfo
  {publisher} {Butterworth - Heinemann, Oxford},\ \bibinfo {year}
  {1984})\BibitemShut {NoStop}%
\bibitem [{\citenamefont {Boyd}(2008)}]{Boyd08}%
  \BibitemOpen
  \bibfield  {author} {\bibinfo {author} {\bibfnamefont {R.~W.}\ \bibnamefont
  {Boyd}},\ }\href@noop {} {\emph {\bibinfo {title} {Nonlinear Optics}}}\
  (\bibinfo  {publisher} {Academ. Press, Rochester, N.Y.},\ \bibinfo {year}
  {2008})\BibitemShut {NoStop}%
\bibitem [{\citenamefont {Krutyanskiy}\ \emph {et~al.}(2013)\citenamefont
  {Krutyanskiy}, \citenamefont {Kolmychek}, \citenamefont {Gribkov},
  \citenamefont {Karashtin}, \citenamefont {Skorohodov},\ and\ \citenamefont
  {Murzina}}]{Krutyanskiy13}%
  \BibitemOpen
  \bibfield  {author} {\bibinfo {author} {\bibfnamefont {V.~L.}\ \bibnamefont
  {Krutyanskiy}}, \bibinfo {author} {\bibfnamefont {I.~A.}\ \bibnamefont
  {Kolmychek}}, \bibinfo {author} {\bibfnamefont {B.~A.}\ \bibnamefont
  {Gribkov}}, \bibinfo {author} {\bibfnamefont {E.~A.}\ \bibnamefont
  {Karashtin}}, \bibinfo {author} {\bibfnamefont {E.~V.}\ \bibnamefont
  {Skorohodov}}, \ and\ \bibinfo {author} {\bibfnamefont {T.~V.}\ \bibnamefont
  {Murzina}},\ }\href {\doibase 10.1103/PhysRevB.88.094424} {\bibfield
  {journal} {\bibinfo  {journal} {Phys. Rev. B}\ }\textbf {\bibinfo {volume}
  {88}},\ \bibinfo {pages} {094424} (\bibinfo {year} {2013})}\BibitemShut
  {NoStop}%
\bibitem [{\citenamefont {Kolmychek}\ \emph {et~al.}(2015)\citenamefont
  {Kolmychek}, \citenamefont {Krutyanskiy}, \citenamefont {Murzina},
  \citenamefont {Sapozhnikov}, \citenamefont {Karashtin}, \citenamefont
  {Rogov},\ and\ \citenamefont {Fraerman}}]{Kolmychek15}%
  \BibitemOpen
  \bibfield  {author} {\bibinfo {author} {\bibfnamefont {I.~A.}\ \bibnamefont
  {Kolmychek}}, \bibinfo {author} {\bibfnamefont {V.~L.}\ \bibnamefont
  {Krutyanskiy}}, \bibinfo {author} {\bibfnamefont {T.~V.}\ \bibnamefont
  {Murzina}}, \bibinfo {author} {\bibfnamefont {M.~V.}\ \bibnamefont
  {Sapozhnikov}}, \bibinfo {author} {\bibfnamefont {E.~A.}\ \bibnamefont
  {Karashtin}}, \bibinfo {author} {\bibfnamefont {V.~V.}\ \bibnamefont
  {Rogov}}, \ and\ \bibinfo {author} {\bibfnamefont {A.~A.}\ \bibnamefont
  {Fraerman}},\ }\href {\doibase 10.1364/JOSAB.32.000331} {\bibfield  {journal}
  {\bibinfo  {journal} {J. Opt. Soc. Am. B}\ }\textbf {\bibinfo {volume}
  {32}},\ \bibinfo {pages} {331} (\bibinfo {year} {2015})}\BibitemShut
  {NoStop}%
\bibitem [{\citenamefont {Karashtin}\ and\ \citenamefont
  {Fraerman}(2018)}]{Karashtin18}%
  \BibitemOpen
  \bibfield  {author} {\bibinfo {author} {\bibfnamefont {E.~A.}\ \bibnamefont
  {Karashtin}}\ and\ \bibinfo {author} {\bibfnamefont {A.~A.}\ \bibnamefont
  {Fraerman}},\ }\href {\doibase 10.1088/1361-648X/aab56d} {\bibfield
  {journal} {\bibinfo  {journal} {J. Phys. Condens. Matter}\ }\textbf {\bibinfo
  {volume} {30}},\ \bibinfo {pages} {165801} (\bibinfo {year}
  {2018})}\BibitemShut {NoStop}%
\bibitem [{\citenamefont {Rzhevsky}\ \emph {et~al.}(2007)\citenamefont
  {Rzhevsky}, \citenamefont {Krichevtsov}, \citenamefont {B\"urgler},\ and\
  \citenamefont {Schneider}}]{Rzhevsky07}%
  \BibitemOpen
  \bibfield  {author} {\bibinfo {author} {\bibfnamefont {A.~A.}\ \bibnamefont
  {Rzhevsky}}, \bibinfo {author} {\bibfnamefont {B.~B.}\ \bibnamefont
  {Krichevtsov}}, \bibinfo {author} {\bibfnamefont {D.~E.}\ \bibnamefont
  {B\"urgler}}, \ and\ \bibinfo {author} {\bibfnamefont {C.~M.}\ \bibnamefont
  {Schneider}},\ }\href {\doibase 10.1103/PhysRevB.75.144416} {\bibfield
  {journal} {\bibinfo  {journal} {Phys. Rev. B}\ }\textbf {\bibinfo {volume}
  {75}},\ \bibinfo {pages} {144416} (\bibinfo {year} {2007})}\BibitemShut
  {NoStop}%
\bibitem [{\citenamefont {Shahbazi}\ \emph
  {et~al.}(2018{\natexlab{a}})\citenamefont {Shahbazi}, \citenamefont {Kim},
  \citenamefont {Nembach}, \citenamefont {Shaw}, \citenamefont {Bischof},
  \citenamefont {Rossell}, \citenamefont {Jeudy}, \citenamefont {Moore},\ and\
  \citenamefont {Marrows}}]{Shahbazi18}%
  \BibitemOpen
  \bibfield  {author} {\bibinfo {author} {\bibfnamefont {K.}~\bibnamefont
  {Shahbazi}}, \bibinfo {author} {\bibfnamefont {J.-V.}\ \bibnamefont {Kim}},
  \bibinfo {author} {\bibfnamefont {H.~T.}\ \bibnamefont {Nembach}}, \bibinfo
  {author} {\bibfnamefont {J.~M.}\ \bibnamefont {Shaw}}, \bibinfo {author}
  {\bibfnamefont {A.}~\bibnamefont {Bischof}}, \bibinfo {author} {\bibfnamefont
  {M.~D.}\ \bibnamefont {Rossell}}, \bibinfo {author} {\bibfnamefont
  {V.}~\bibnamefont {Jeudy}}, \bibinfo {author} {\bibfnamefont {T.~A.}\
  \bibnamefont {Moore}}, \ and\ \bibinfo {author} {\bibfnamefont {C.~H.}\
  \bibnamefont {Marrows}},\ }\href@noop {} {\  (\bibinfo {year}
  {2018}{\natexlab{a}})},\ \Eprint {http://arxiv.org/abs/1810.03454}
  {arXiv:1810.03454 [cond-mat]} \BibitemShut {NoStop}%
\bibitem [{\citenamefont {Shahbazi}\ \emph
  {et~al.}(2018{\natexlab{b}})\citenamefont {Shahbazi}, \citenamefont {Hrabec},
  \citenamefont {Moretti}, \citenamefont {Ward}, \citenamefont {Moore},
  \citenamefont {Jeudy}, \citenamefont {Martinez},\ and\ \citenamefont
  {Marrows}}]{Shahbazi18_2}%
  \BibitemOpen
  \bibfield  {author} {\bibinfo {author} {\bibfnamefont {K.}~\bibnamefont
  {Shahbazi}}, \bibinfo {author} {\bibfnamefont {A.}~\bibnamefont {Hrabec}},
  \bibinfo {author} {\bibfnamefont {S.}~\bibnamefont {Moretti}}, \bibinfo
  {author} {\bibfnamefont {M.~B.}\ \bibnamefont {Ward}}, \bibinfo {author}
  {\bibfnamefont {T.~A.}\ \bibnamefont {Moore}}, \bibinfo {author}
  {\bibfnamefont {V.}~\bibnamefont {Jeudy}}, \bibinfo {author} {\bibfnamefont
  {E.}~\bibnamefont {Martinez}}, \ and\ \bibinfo {author} {\bibfnamefont
  {C.~H.}\ \bibnamefont {Marrows}},\ }\href@noop {} {\  (\bibinfo {year}
  {2018}{\natexlab{b}})},\ \Eprint {http://arxiv.org/abs/1809.03217}
  {arXiv:1809.03217 [cond-mat]} \BibitemShut {NoStop}%
\bibitem [{\citenamefont {Wilts}\ and\ \citenamefont
  {Humphrey}(1968)}]{Wilts68}%
  \BibitemOpen
  \bibfield  {author} {\bibinfo {author} {\bibfnamefont {C.~H.}\ \bibnamefont
  {Wilts}}\ and\ \bibinfo {author} {\bibfnamefont {F.~B.}\ \bibnamefont
  {Humphrey}},\ }\href {\doibase 10.1063/1.1656219} {\bibfield  {journal}
  {\bibinfo  {journal} {Journal of Applied Physics}\ }\textbf {\bibinfo
  {volume} {39}},\ \bibinfo {pages} {1191} (\bibinfo {year}
  {1968})}\BibitemShut {NoStop}%
\bibitem [{\citenamefont {Andr\"a}\ \emph {et~al.}(1960)\citenamefont
  {Andr\"a}, \citenamefont {M\'alek}, \citenamefont {Sch\"uppel},\ and\
  \citenamefont {Stemme}}]{Andr60}%
  \BibitemOpen
  \bibfield  {author} {\bibinfo {author} {\bibfnamefont {W.}~\bibnamefont
  {Andr\"a}}, \bibinfo {author} {\bibfnamefont {Z.}~\bibnamefont {M\'alek}},
  \bibinfo {author} {\bibfnamefont {W.}~\bibnamefont {Sch\"uppel}}, \ and\
  \bibinfo {author} {\bibfnamefont {O.}~\bibnamefont {Stemme}},\ }\href@noop {}
  {\bibfield  {journal} {\bibinfo  {journal} {Journal of Applied Physics}\
  }\textbf {\bibinfo {volume} {31}},\ \bibinfo {pages} {442} (\bibinfo {year}
  {1960})}\BibitemShut {NoStop}%
\bibitem [{\citenamefont {Belmeguenai}\ \emph {et~al.}(2015)\citenamefont
  {Belmeguenai}, \citenamefont {Adam}, \citenamefont {Roussign\'e},
  \citenamefont {Eimer}, \citenamefont {Devolder}, \citenamefont {Kim},
  \citenamefont {Cherif}, \citenamefont {Stashkevich},\ and\ \citenamefont
  {Thiaville}}]{Belmeguenai15}%
  \BibitemOpen
  \bibfield  {author} {\bibinfo {author} {\bibfnamefont {M.}~\bibnamefont
  {Belmeguenai}}, \bibinfo {author} {\bibfnamefont {J.-P.}\ \bibnamefont
  {Adam}}, \bibinfo {author} {\bibfnamefont {Y.}~\bibnamefont {Roussign\'e}},
  \bibinfo {author} {\bibfnamefont {S.}~\bibnamefont {Eimer}}, \bibinfo
  {author} {\bibfnamefont {T.}~\bibnamefont {Devolder}}, \bibinfo {author}
  {\bibfnamefont {J.-V.}\ \bibnamefont {Kim}}, \bibinfo {author} {\bibfnamefont
  {S.~M.}\ \bibnamefont {Cherif}}, \bibinfo {author} {\bibfnamefont
  {A.}~\bibnamefont {Stashkevich}}, \ and\ \bibinfo {author} {\bibfnamefont
  {A.}~\bibnamefont {Thiaville}},\ }\href {\doibase 10.1103/PhysRevB.91.180405}
  {\bibfield  {journal} {\bibinfo  {journal} {Phys. Rev. B}\ }\textbf {\bibinfo
  {volume} {91}},\ \bibinfo {pages} {180405} (\bibinfo {year}
  {2015})}\BibitemShut {NoStop}%
\bibitem [{\citenamefont {Woo}\ \emph {et~al.}(2014)\citenamefont {Woo},
  \citenamefont {Mann}, \citenamefont {Tan}, \citenamefont {Caretta},\ and\
  \citenamefont {Beach}}]{Woo14}%
  \BibitemOpen
  \bibfield  {author} {\bibinfo {author} {\bibfnamefont {S.}~\bibnamefont
  {Woo}}, \bibinfo {author} {\bibfnamefont {M.}~\bibnamefont {Mann}}, \bibinfo
  {author} {\bibfnamefont {A.~J.}\ \bibnamefont {Tan}}, \bibinfo {author}
  {\bibfnamefont {L.}~\bibnamefont {Caretta}}, \ and\ \bibinfo {author}
  {\bibfnamefont {G.~S.~D.}\ \bibnamefont {Beach}},\ }\href {\doibase
  10.1063/1.4902529} {\bibfield  {journal} {\bibinfo  {journal} {Applied
  Physics Letters}\ }\textbf {\bibinfo {volume} {105}},\ \bibinfo {pages}
  {212404} (\bibinfo {year} {2014})}\BibitemShut {NoStop}%
\bibitem [{\citenamefont {He}\ \emph {et~al.}(2018)\citenamefont {He},
  \citenamefont {Li}, \citenamefont {Zhu}, \citenamefont {Zhang}, \citenamefont
  {Peng}, \citenamefont {Li}, \citenamefont {Li}, \citenamefont {Wei},
  \citenamefont {Zhao}, \citenamefont {Zhang}, \citenamefont {Wang},
  \citenamefont {Lin}, \citenamefont {Gu}, \citenamefont {Yu}, \citenamefont
  {Cai},\ and\ \citenamefont {Shen}}]{Min18}%
  \BibitemOpen
  \bibfield  {author} {\bibinfo {author} {\bibfnamefont {M.}~\bibnamefont
  {He}}, \bibinfo {author} {\bibfnamefont {G.}~\bibnamefont {Li}}, \bibinfo
  {author} {\bibfnamefont {Z.}~\bibnamefont {Zhu}}, \bibinfo {author}
  {\bibfnamefont {Y.}~\bibnamefont {Zhang}}, \bibinfo {author} {\bibfnamefont
  {L.}~\bibnamefont {Peng}}, \bibinfo {author} {\bibfnamefont {R.}~\bibnamefont
  {Li}}, \bibinfo {author} {\bibfnamefont {J.}~\bibnamefont {Li}}, \bibinfo
  {author} {\bibfnamefont {H.}~\bibnamefont {Wei}}, \bibinfo {author}
  {\bibfnamefont {T.}~\bibnamefont {Zhao}}, \bibinfo {author} {\bibfnamefont
  {X.-G.}\ \bibnamefont {Zhang}}, \bibinfo {author} {\bibfnamefont
  {S.}~\bibnamefont {Wang}}, \bibinfo {author} {\bibfnamefont {S.-Z.}\
  \bibnamefont {Lin}}, \bibinfo {author} {\bibfnamefont {L.}~\bibnamefont
  {Gu}}, \bibinfo {author} {\bibfnamefont {G.}~\bibnamefont {Yu}}, \bibinfo
  {author} {\bibfnamefont {J.~W.}\ \bibnamefont {Cai}}, \ and\ \bibinfo
  {author} {\bibfnamefont {B.-g.}\ \bibnamefont {Shen}},\ }\href {\doibase
  10.1103/PhysRevB.97.174419} {\bibfield  {journal} {\bibinfo  {journal} {Phys.
  Rev. B}\ }\textbf {\bibinfo {volume} {97}},\ \bibinfo {pages} {174419}
  (\bibinfo {year} {2018})}\BibitemShut {NoStop}%
\end{thebibliography}%

\end{document}